\documentclass[10pt]{raa}            
\usepackage[T1]{fontenc}
\usepackage{ae,aecompl}
\usepackage{graphicx,times}             
\begin{document}

   \title{The galaxy luminosity function in the LAMOST Complete Spectroscopic
Survey of Pointing Area at the Southern Galactic Cap
}

   \volnopage{Vol.0 (20xx) No.0, 000--000}      
   \setcounter{page}{1}          

   \author{Pin-Song\ Zhao
      \inst{1,2}
   \and Hong\ Wu
      \inst{1,2}
   \and C.\ K.\ Xu
      \inst{3}
   \and Ming\ Yang
      \inst{4}
   \and Fan\ Yang
      \inst{1}
   \and Yi-Nan\ Zhu
      \inst{1}
   \and Man\ I\ Lam
      \inst{5}
   \and Jun-Jie\ Jin
      \inst{1,2}
   \and Hai-Long\ Yuan
      \inst{1}
   \and Hao-Tong\ Zhang
      \inst{1}
   \and Shi-Yin\ Shen
      \inst{5}
   \and Jian-Rong\ Shi
      \inst{1}
   \and A-Li\ Luo
      \inst{1}
   \and Xue-Bing\ Wu
      \inst{6}
   \and Yong-Heng\ Zhao
      \inst{1}
   \and Yi-Peng\ Jing
      \inst{7}
   }

   \institute{Key Laboratory of Optical Astronomy, National Astronomical Observatories,
      Chinese Academy of Sciences,
      20A Datun Road, Chaoyang District, Beijing 100101, China; {\it zps@bao.ac.cn}\\
        \and
             School of Astronomy and Space Science, University of Chinese Academy of Sciences,
      Beijing 100049, China\\
        \and
             Chinese Academy of Sciences South America Center for Astronomy, China-Chile Joint
      Center for Astronomy, 
      Camino El Observatorio 1515, Las Condes, Santiago, Chile\\
        \and
             IAASARS, National Observatory of Athens, Vas. Pavlou \& I. Metaxa, Penteli 15236, Greece\\
        \and
             Key Laboratory for Research in Galaxies and Cosmology, Shanghai Astronomical Observatory,
             Chinese Academy of Sciences, 80 Nandan Road, Shanghai 200030, China\\
        \and
             Department of Astronomy, School of Physics, Peking University, Beijing 100871, China\\
        \and
             School of Physics and Astronomy, Shanghai Jiao Tong University, 800 Dongchuan Road,
Shanghai 200240, China\\
   }

   \date{xxx}

\abstract{
  We present optical luminosity functions (LFs) of galaxies in the
  $^{0.1}g$, $^{0.1}r$, and $^{0.1}i$ bands, calculated using data in
  $\sim 40$ $deg^{2}$ sky area of LAMOST Complete Spectroscopic Survey
  of Pointing Area (LaCoSSPAr) in Southern Galactic Cap.  Redshifts
  for galaxies brighter $r = 18.1$ were obtained mainly with LAMOST.
  In each band, LFs derived using both parametric and
  non-parametric maximum likelihood methods agree well with each
  other.  In the $^{0.1}r$ band, our fitting parameters of the
  Schechter function are
  $\phi_{*}=(1.65\pm0.36)\times10^{-2}h^{3}Mpc^{-3}$,
  $M_{*}=-20.69\pm0.06$ mag, and $\alpha=-1.12\pm0.08$, in agreements
  with previous studies. Separate LFs are also derived for emission
  line galaxies and absorption line galaxies, respectively.  The LFs of
  absorption line galaxies show a dip at $^{0.1}r \sim 18.5$ and can
  be well fitted by a double-Gaussian function, suggesting
  a bi-modality in passive galaxies.
}

   \authorrunning{Zhao et al. }            
   \titlerunning{LoCaSSPAr luminosity functions }  

   \maketitle

%
%
\section{Introduction}           
\label{sect:intro}

Luminosity is one of the most basic properties of galaxies.  Studies 
of galaxy luminosity functions (LFs) give direct statistical estimates for 
the space density of galaxies with respect to their luminosities, and provide
important information about the galaxy formation and evolution.

In recent years, many large spectroscopic surveys have been conducted to
investigate the nearby universe.  Among them, the Center for Astrophysics
(CfA) Survey (Huchra et al.~\cite{1983apjs...52..89}), the Two-Degree
Field (2dF) Galaxy Redshift Survey (Lewis et
al.~\cite{2002mnras...334..673}), the Sloan Digital Sky Survey (SDSS;
York et al.~\cite{2000aj...120..1579}), and the Galaxy And Mass
Assembly (GAMA) redshift survey (Driver et al.~\cite{Driver2009};
Baldry et al.~\cite{2010mnras...404..86}), etc., were all very
successful and gave us a better understanding of the universe.  Thanks
to these surveys, many investigations of galaxy LFs
have been done, providing important observing constraints on theories of
galaxy formation and evolution.

Blanton et
al.\ (\cite{2001aj...121..2358}) calculated the galaxy LFs in SDSS $ugriz$ bands
using SDSS commissioning data and discussed the dependence
of luminosity on surface brightness, color and morphology. 
Blanton et al.\ (\cite{2003apj...592..819B}) fitted the LFs
using two parameters, Q and P, to study effects of luminosity and
density evolution, respectively. Montero-Dorta \& Prada\
(\cite{2009mnras...399..1106M}) calculated the luminosity function
with a sample selected from 
 SDSS DR6 (Adelman-McCarthy et al.~\cite{Adelman-McCarthy2008}),
and found a remarkable excess at the bright end of the
$^{0.1}u$ band LF. Loveday et al.\ (\cite{2012mnras...420..1239L})
focused on the evolution of the LFs in a redshift
range of $0.002<z<0.5$ and pointed out different evolution features
between blue galaxies and red galaxies based on GAMA core data release
(Driver et al.~\cite{Driver2011}).

At higher redshift ($z>0.5$), Willmer et al.\ (\cite{2006Willmer})
constructed $B$-band LFs of red and blue galaxies in
different redshift slices from $z\sim0.2$ to $z\sim1.2$ based on the
Deep Evolutionary Exploratory Probe 2 redshift survey (DEEP2; Davis et
al.~\cite{Davis2003}), and found a more significant
luminosity evolution for blue
galaxies while for red galaxies a more significant density evolution.
Montero-Dorta et al.\ (\cite{Montero-Dorta2016}) used the Baryon
Oscillation Spectroscopic Survey (BOSS; Dawson et
al.~\cite{Dawson2013}) high redshift sample, and computed the high
mass end of the SDSS $^{0.55}i$ band luminosity functions of red
sequence galaxies at redshift $z\sim0.55$, they suggested that these
red sequence galaxies formed at redshift $z=1.5-3$.
L$\acute{o}$pez-Sanjuan et al.\ (\cite{Lopez2017}) studied the B-band
luminosity functions for star-forming and quiescent galaxies based on
ALHAMBRA (Advanced, Large, Homogeneous Area, Mediam-Band Redshift
Astronomical) survey (Moles et al.~\cite{Moles2008}), and provided a
distinct understanding of the evolution of B-band luminosity function
and luminosity density for different types of galaxies since $z\sim 1$.

The Large Sky Area Multi-Object Fiber Spectroscopic Telescope (LAMOST)
is a Wang-Su reflecting Schmidt telescope (Wang et
al.~\cite{1996apopt...35..5155}; Su \& Cui~\cite{2004chjaa...4..1};
Cui et al.~\cite{2012raa...12..1197}; Zhao et
al.~\cite{2012raa...12..723}) located in Xinglong Station of National
Astronomical Observatory, Chinese Academy of Sciences (NAOC).  Thanks
to its $\sim 20$ $deg^{2}$ field of view (FOV) and 4000 fibers,
LAMOST can spectroscopically observe more than 3000 scientific targets
simultaneously (nearly 5 times more than SDSS), making it efficient to
obtain spectra of celestial objects.  LAMOST ExtraGAlactic Survey
(LEGAS), an important part of LAMOST scientific survey, aims to
cover $\sim8000 deg^{2}$ of the Northern Galactic Cap (NGC) and
$\sim3500 deg^2$ of the Southern Galactic Cap (SGC) and take hundreds
of thousands of spectra for extra-galactic objects with redshifts  
$z<0.3$ in five years (Yang et al.~\cite{2018yang}).  When
finishing its extra-galactic survey, LAMOST will provide a catalog,
covering large sky area ($\sim 11500 deg^{2}$ in total) and containing millions of spectroscopic information of galaxies.  

This work is based on LAMOST Complete Spectroscopic
Survey of Pointing Area (LaCoSSPAr), an early project of LEGAS. The
LaCoSSPAr is a LAMOST key project aiming at observing all sources
(galactic and extra-galactic) with a magnitude limit of $14.0<r<18.1$
in selected two 20 $degree^{2}$ regions in SGC, where the faint
magnitude limit is 0.1 mag deeper than LAMOST designed and
0.33 mag deeper than that of SDSS legacy survey.  This survey is
designed to investigate the completeness and selection effects in the
wider LEGAS survey (Yang et al.~\cite{2018yang}).  By using the
spectra observed by LAMOST and cross-matching with data of other
photometric surveys, the galaxy LFs can be
investigated in specific bands.  Our fields locate in the SGC, where
the footprint covered by the SDSS is small.  Meanwhile, thanks to the
high galactic latitude, our galaxy sample suffers less from the effects of
Galactic extinction.

In this paper, we use the galaxy redshift sample based on LaCoSSPAr,
which is the most complete sample in LEGAS up to now, and combine
with SDSS Petrosian magnitudes, to estimate the galaxy LFs 
in SDSS $^{0.1}g$, $^{0.1}r$, $^{0.1}i$ bands.  Our 
sample has a fainter limiting magnitude than SDSS and
our goal is to achieve a better understanding about the faint end of
galaxy LFs.  In section 2, we give an introduction to LAMOST data and
data reduction, and describe our sample selection and correction for
the incompleteness.  In section 3, we introduce the methods used to
estimate the galaxy LFs.  And in section 4, we present
the results of the LFs and discussion.  A summary is
presented in section 5.  Throughout this paper, we adopt a
Friedmann-Robertson-Walker cosmological world model with constants of
$\Omega_{m}=0.3$,$\Omega_{\Lambda}=0.7$ and $H_{0}=70$
$kms^{-1}Mpc^{-1}$.


\section{Sample} \label{sec:sample}
\subsection{LaCoSSPAr, data and data reduction} \label{subsec:data} 

The LaCoSSPAr surveys two $\sim20$ $degree^{2}$ regions in SGC
  with limiting magnitude of $r=18.1$ mag.  Originally, the plan was to
  select a higher density region and a lower density region to test
  possible environmental effects.  The high
  density field (Field B: $R.A.=21.53^{\circ}$,
  $Dec.=-2.20^{\circ}$) is selected to cover a large Abell rich 
  cluster (Abell et al.~\cite{Abell1989}), and the
  low density field (Field A: $R.A.=37.88^{\circ}$,
  $Dec.=3.44^{\circ}$) is selected in a blank region near
  Field B (as shown in Figure 1 in Yang et
  al.~\cite{2018yang}).  However, it was found later that Field A 
  (low density field) actually contains 11 faint Abell and Zwicky clusters
  and therefore may not represent low density regions. The effects
  of the field selection will be discussed in Section 4.

The input catalog for targets of LaCoSSPAr survey were selected
  from the Data Release 9 (DR9; Ahn et al.~\cite{Ahn2012}) of the SDSS
  PhotoPrimary database, using the criteria of $14.0<r<18.1$ and type
  =`Galaxy'. Sources (936) were excluded when they are in the
  following special regions that are not observed by LAMOST: (1) in
  the LAMOST five guide CCDs fields, (2) within 10" from bright stars,
  and (3) in dense regions. The final LaCoSSPAr target catalog
  contains 5623 sources, of which 5442 (96.8\%) were observed
  successfully, and 181 (3.2\%) failed mainly due to bad fibers. 

The raw data of successful observations
were first reduced by the LAMOST 2D and 1D pipelines (Luo
et al.~\cite{2012raa...12..1243}), which include bias subtraction,
flat-fielding through twilight exposures, cosmic-ray removal, spectrum
extraction, wavelength calibration, sky subtraction and exposure
coaddition.  However, for many spectra with relatively low SNR
  the pipeline does not work well.  Low SNR makes it hard to recognize
  diagnostic lines. Also, bad sky line subtraction often introduces
  fake lines that affect significantly the redshift measurement.
  Consequently, only about a third of total observed galaxies obtained
  redshifts from the pipeline.  In order to achieve better redshift
detection rate, additional data processing of
the 1D spectrum is carried out using our own software (Yang et
al.~\cite{2018yang}).  Briefly speaking, in order to improve the
results of sky line subtraction, in the residual spectrum we replaced
all $>3\sigma$ points around each skyline ($\pm 15 \AA$) by the values
of continuum fitting.  After this, we inspected each spectrum by eyes
(by at least two individuals) and re-measured the redshift by
identifying emission lines and absorption lines.  These new steps
improved significantly the success rate of redshift detection (Bai et
al.~\cite{bai2017}). 
Redshifts of 3098 sources were detected, corresponding to a detection rate of 55\%. They have a  median redshift of 
$\bar{z}=0.104$ and typical uncertainty of $\sigma_{z}/(1+z)<0.001$.

\subsection{Parent Sample and Redshift Completeness} \label{subsec:incomple}

The parent sample for the LF calculations is based on the
  LaCoSSPAr target catalog (see Section 2.1). Actually, many sources
  in that catalog are stars or fake targets identified mistakenly as
  galaxies by SDSS. In order to exclude them, we inspected the images
  of all sources visually on the SDSS navigator tool and discarded
  those showing obvious characteristics of a star, or no show at all
  (fake sources). Furthermore, among the 3098 sources with LAMOST
  redshifts, 60 were found to be Galactic sources with z=0 and
  therefore were excluded.  Finally, our parent sample contains 5531
  galaxies, of which 3038 having redshifts from LAMOST. In addition,
  457 galaxies in the sample have SDSS redshifts but no LAMOST
  redshifts. Altogether, 3495 galaxies in our sample have measured
  redshifts, corresponding to a redshift completeness of 63\%. For
  galaxies brighter than the magnitude limit of SDSS spectroscopic
  survey, r=17.77 mag, the redshift completeness of the sample is 69\%
  (2592/3749).

\begin{figure*}
\centering

\includegraphics[bb=450 0 900 450,width=0.35\textwidth]{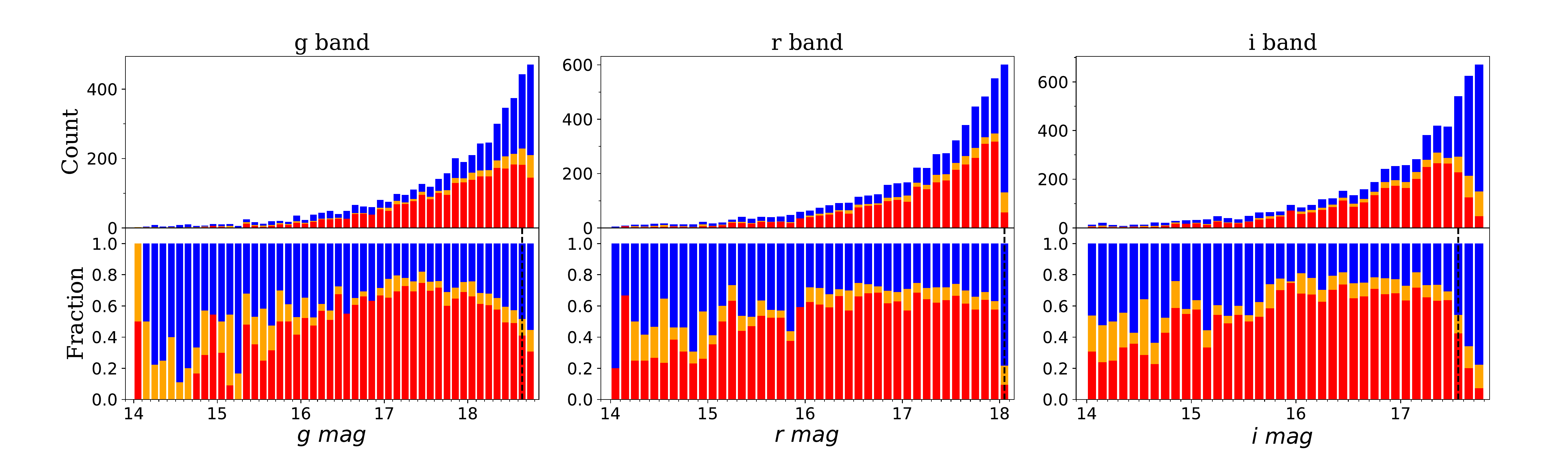}
\caption {Upper panel: the histogram of Petrosian
    magnitude in each band for all galaxies in the parent
  sample. Different colors represent for galaxies with redshifts from
  LaCoSSPAr (red), galaxies having no LaCoSSPAr redshifts but having
  redshifts from SDSS (orange) and galaxies having no redshifts
  (blue).  Lower panel: the fraction for different classes within each
  magnitude bin.  The black dashed lines show the upper magnitude
  limit of subsample.  }

\label{fig:completeness}
\end{figure*}

In Figure \ref{fig:completeness}, the magnitude
distribution of galaxies in the parent sample is presented.
For each galaxy,
the Galactic extinction was corrected using the
dust maps of Schlegel, Finkbeiner \& Davis\
(\cite{1998apj...500..525S}).    
Upper panel shows
histograms of magnitudes in $gri$ bands for all visually-examined
photometric galaxies.  We use different colors to represent galaxies
with redshifts from LaCoSSPAr (red), galaxies with redshifts from SDSS
(orange) and galaxies having no redshifts (blue) in each bin.  Lower
panel gives the fraction for different classes within each magnitude
bin.  The black dashed lines mark the magnitude limits of 
corresponding subsamples used in the calculation of individual
LFs. Beyond these limits, the completeness (i.e. the ratio between
galaxies with redshifts and galaxies identified photometrically)
drops rapidly below 50\%.
Figure \ref{fig:completeness} shows that the completeness of faint
galaxies is better than that of bright galaxies. This
counter-intuitive result deserves some explanation. It appears that
the success of redshift detection depends sensitively on how accurately
the fiber position coincides with the target position.  Because
targets fainter than $r=16$ were observed with longer integration times
and more repeats (Yang et al.~\cite{2018yang}), 
they are more resilient against 
the effect of bad fiber position, and therefore have better detection
rates.


\begin{figure*}[!h]
\begin{minipage}{1\linewidth}
\centering
 \includegraphics[bb=150 50 750 450,width=0.5\textwidth]{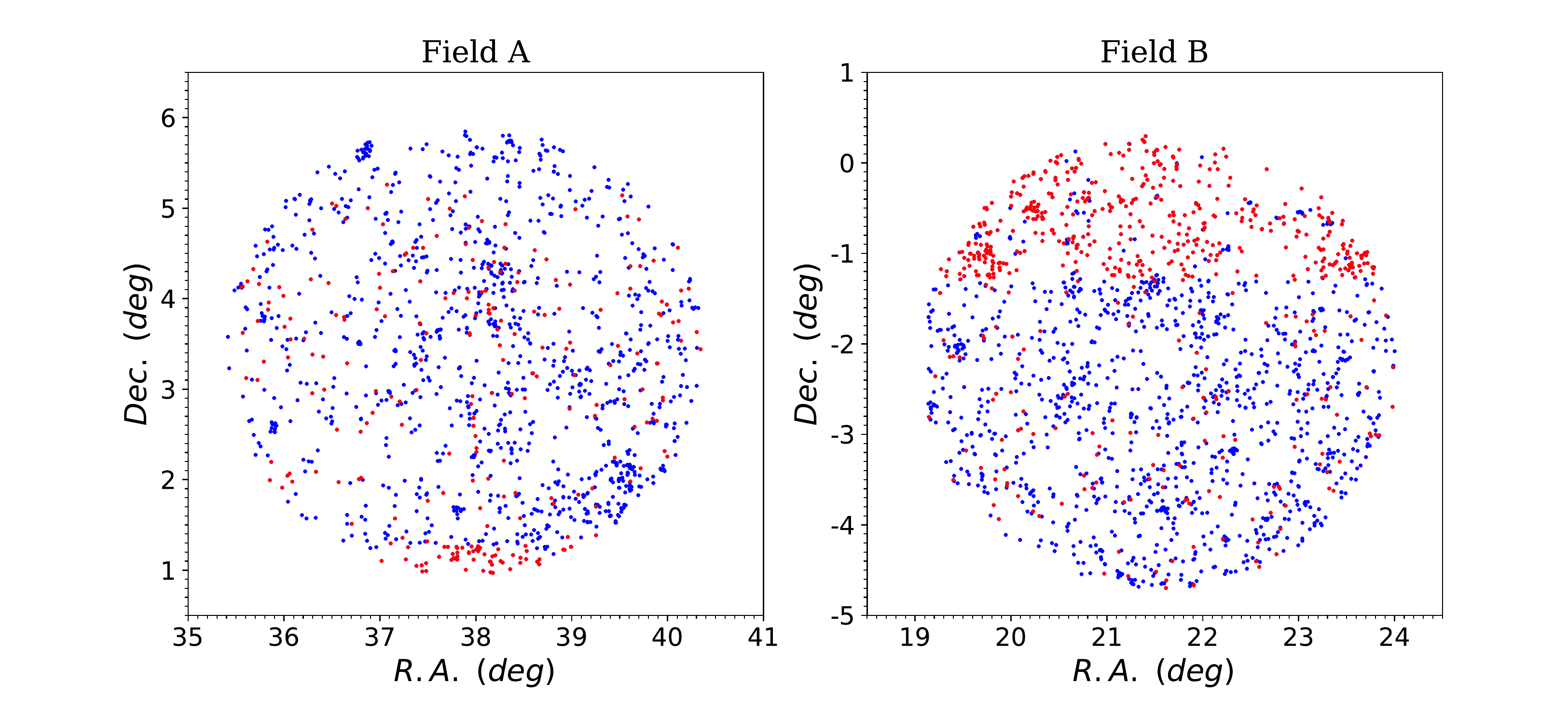}
\caption {Spatial distribution for photometric galaxies (blue dots) and galaxies having SDSS redshifts (red dots) with $r < 17.6$.}
\label{fig:SDSSz_region}
\end{minipage}

\begin{minipage}{1\linewidth}
\centering
 \includegraphics[bb=120 40 750 450,width=0.5\textwidth]{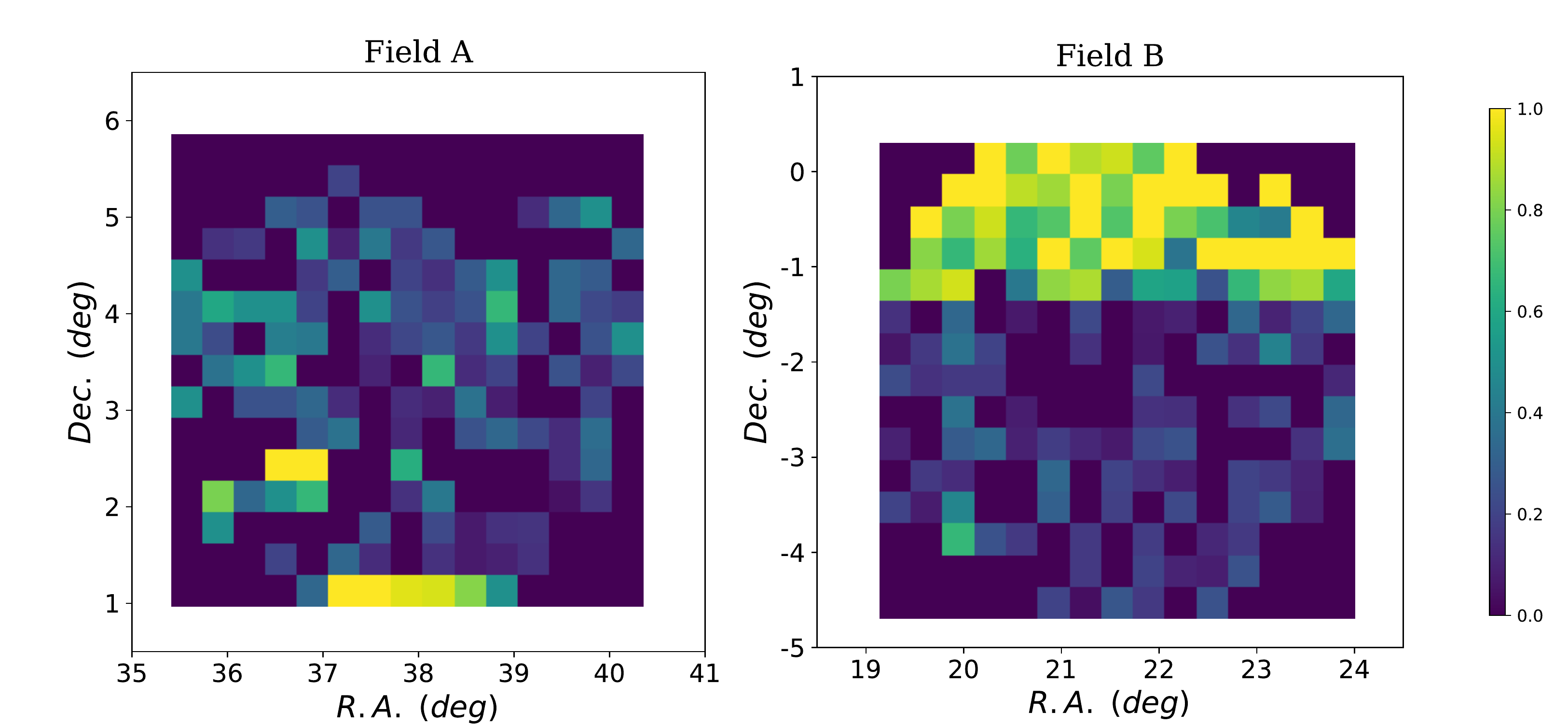}
\caption {The completeness map for SDSS survey in our two fields.
Color bar represents the ratio between galaxies having SDSS redshifts and photometric galaxies for each cell.
}
\label{fig:ratio_region}
\end{minipage}

\end{figure*}

We checked the dependence of LaCoSSPAr redshift incompleteness on
redshift itself by comparing with the SDSS spectroscopic sample.
Given the magnitude limit of SDSS spectroscopic main
galaxy sample, we plot in Figure \ref{fig:SDSSz_region} sky
positions of all photometric galaxies (blue dots) and galaxies with
SDSS redshifts (red dots), both brighter than $r = 17.6$, in our two
fields.  The Stripe 82 of SDSS Legacy Survey (Abazajian et
al.~\cite{Aba2009}) overlaps with our survey, resulting in a higher
SDSS redshift coverage between $-1.25^{\circ}<Dec. < 1.25^{\circ}$, as
presented in Figure \ref{fig:SDSSz_region}.  In order to construct a
complete comparison sample, we divided our two fields into many grid
cells and calculated the ratio between galaxies having SDSS redshifts
and photometric galaxies for each cell.  In Figure
\ref{fig:ratio_region}, we present the completeness map of SDSS
survey in our two fields.  The complete comparison sample 
(here after `sample C') includes all galaxies located within cells
that are $100\%$ complete and with $-1.25^{\circ}<Dec. <
1.25^{\circ}$. It contains 120 galaxies.

\begin{figure*}[!h]
\centering
\includegraphics[bb=0 0 450 450,width=0.45\textwidth]{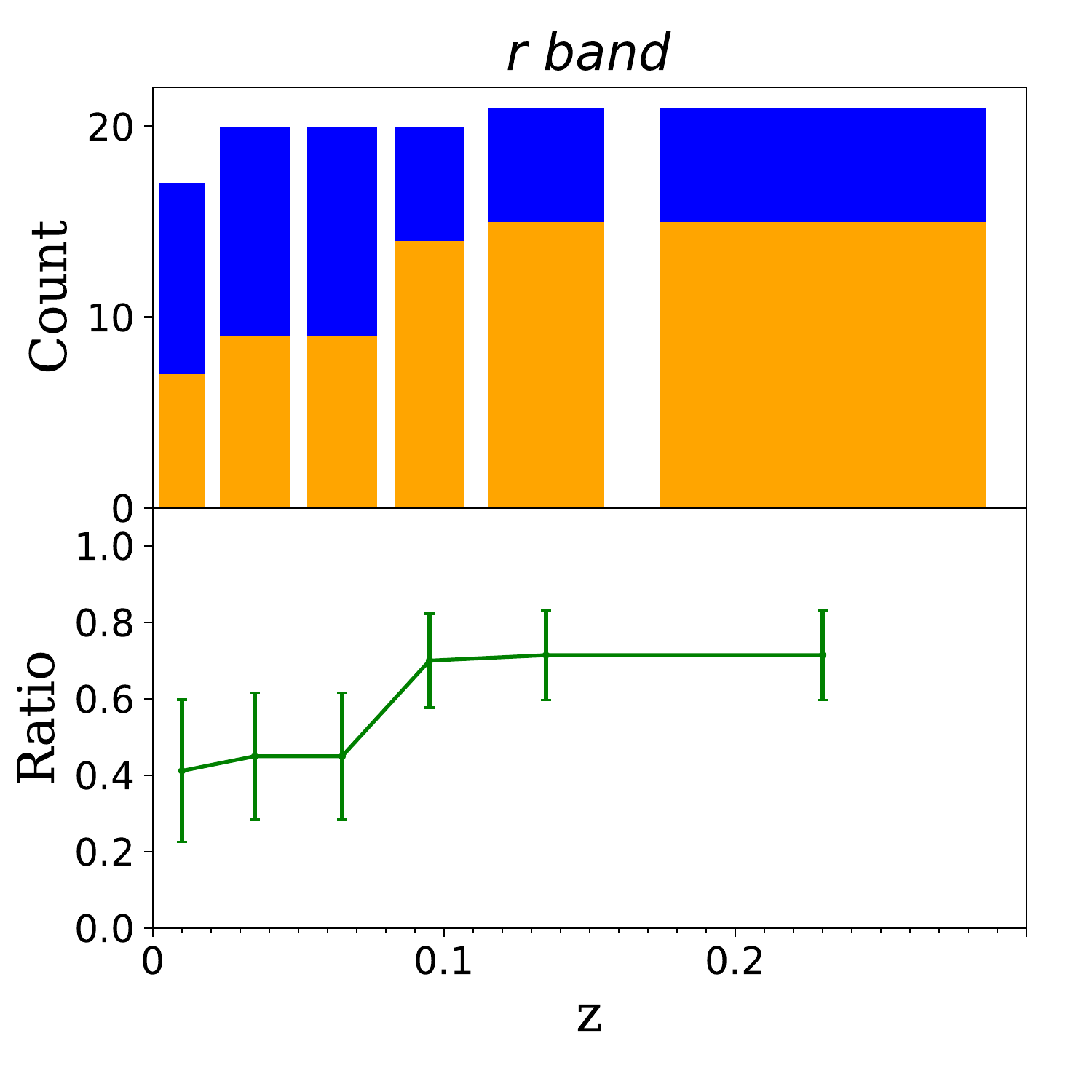}
\caption {Relationship between redshift completeness of LaCoSSPAr and galaxy redshift.
          Upper panel: the histograms of redshift distribution. 
                       The blue bars and orange bars represent counts of redshifts from SDSS survey and from LaCoSSPAr, respectively.
                       We only use the $100\%$ complete cells within $-1.25^{\circ}<Dec. < 1.25^{\circ}$ in Figure \ref{fig:ratio_region} to calculate the galaxy numbers.
                       The bin size are set to ensure roughly equal numbers of galaxies for blue bars in each bin.
          Lower panel: green dots show the ratio of number count of galaxies with LaCoSSPAr redshifts to number count of galaxies with SDSS redshifts in each bin in the upper panel.
                       }
\label{fig:dependz}
\end{figure*}

\begin{figure*}[!h]
\centering
\includegraphics[bb=50 0 950 450,width=0.8\textwidth]{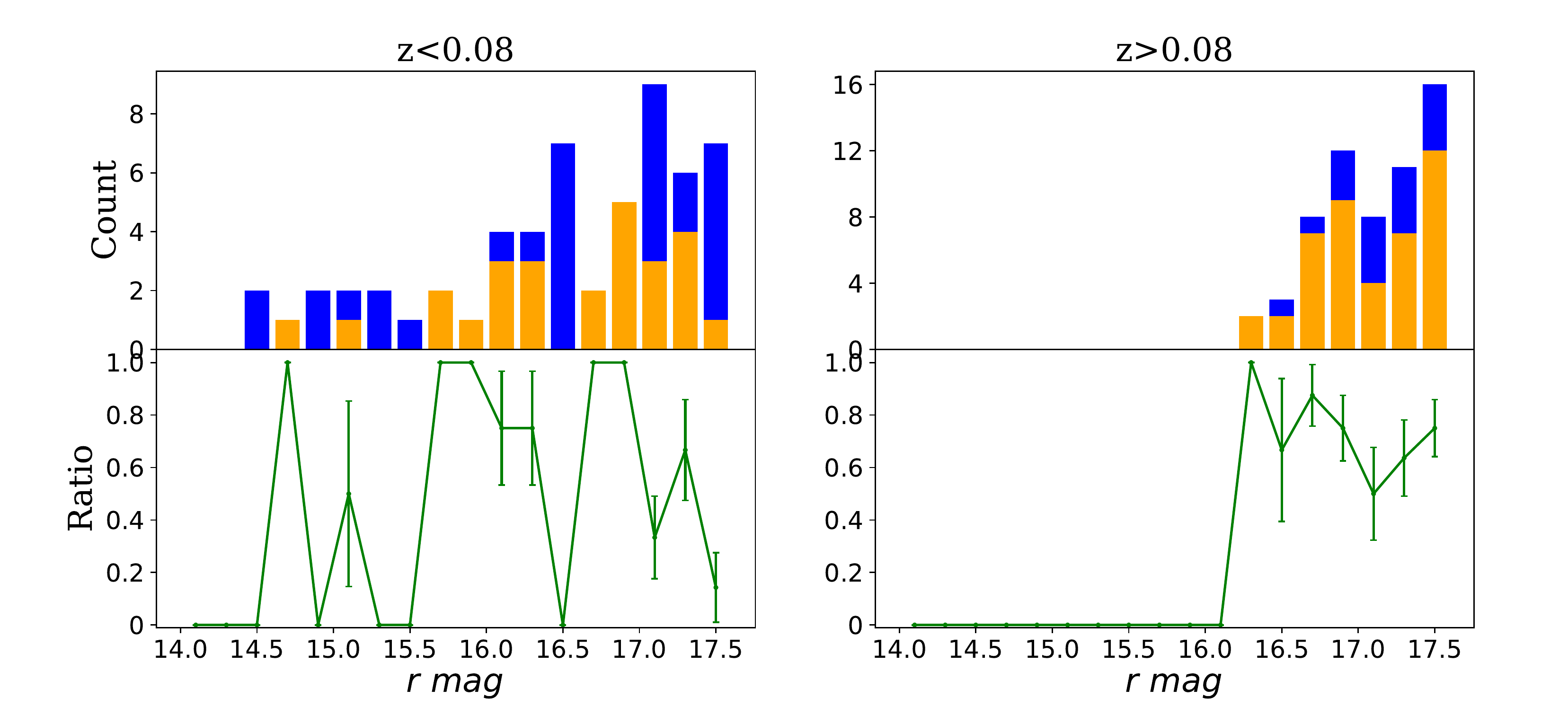}
\caption {Upper panel: the histograms of Petrosian magnitude in r band for galaxies in `sample C' (blue bars) 
                       and galaxies only with redshifts from LaCoSSPAr in `sample C' (orange bars).
                       These galaxies are divided into $z<0.08$ and $z>0.08$ panels.
          Lower panel: the ratio of counts in orange bars to counts in blue bars. 
                       }
\label{fig:120incomp_onmagr}
\end{figure*}

A plot of the redshift dependence of the completeness is presented in
Figure \ref{fig:dependz}.  In the upper panel, histograms of
distributions of SDSS redshifts (blue bars) and LoCaSSPAr redshifts
(orange bars) are plotted for sample C. The bin size has been adjusted
to ensure roughly equal number of galaxies in each bin. The
completeness and error are plotted in the lower panel. It appears
that, for galaxies with $r < 17.6$, the redshift completeness of
LaCoSSPAr has two different levels for $z<\sim 0.08$ and $z>\sim
0.08$: $\sim 0.4$ for low redshift range and $\sim 0.7$ for high
redshift range.  In Figure \ref{fig:120incomp_onmagr}, we
divided the 120 galaxies in `sample C' into a $z>0.08$ subsample and a $z<0.08$
subsample.  Galaxies in the high redshift
subsample are all with $r>16.0$ mag, so they have higher completeness.
In the low redshift subsample, galaxies cover a large magnitude range
from 14.4 mag to 17.6 mag.  Among them, bright galaxies have lower
completeness while faint galaxies still have relative higher
completeness.  It appears that the difference between redshift
incompleteness in two redshift ranges is caused by the different
incompleteness between bright and faint galaxies, as shown in 
Figure \ref{fig:completeness}. 

\subsection{Samples for LFs in different bands} \label{subsec:samselection}

Samples for LFs in different bands were constructed by applying
  corresponding redshift limits and apparent magnitude limits to the
  parent sample.  
The lower magnitude limits were set to be 14.0 mag in all bands.  The
upper magnitude limit in $r$ band was set to be the same with that of
LaCoSSPAr, $r<18.1$ mag.  In other bands, we chose the upper limit at
the magnitude where the redshift completeness falls rapidly (Figure
\ref{fig:completeness}).  As for the redshift limits, we chose the
upper redshift limits at where 98 percent of galaxies are included in
the sample to avoid large noise in the determination of the
normalization at high redshift.  And the lower redshift limits are the
same with that in Blanton et al.\ (\cite{2001aj...121..2358}), which
can reduce the effect of galaxy peculiar velocities when calculating
galaxy luminosity distance.

The lower and upper limits of redshift and magnitude along with the number of galaxies for the samples are listed in Table \ref{tab:table1}.
In this work, we did not include the $u$, $z$ bands because of their relative large photometric uncertainties.
\begin{table}
\begin{center}
 \renewcommand\arraystretch{1.0}
 \caption{Magnitude and redshift limits and galaxy counts of samples for LFs in different bands.}
 \label{tab:table1}
 \begin{tabular}{cccc}
  \hline
  Band & Magnitude limits & Redshift limits & No. of Galaxies \\
  \hline
  $g$ & $14.0<m<18.7$ & $0.016<z<0.23$ & 2718 \\
  $r$ & $14.0<m<18.1$ & $0.016<z<0.27$ & 3412 \\
  $i$ & $14.0<m<17.6$ & $0.016<z<0.275$ & 3235 \\
  \hline
 \end{tabular}
\end{center}
\end{table}

\section{Luminosity functions} \label{sec:LFs}

We used KCORRECT v4\_3 (Blanton et al.~\cite{2007aj...133..734B}) code
to estimate the K-corrections for SDSS magnitudes.  In order to 
compare with LFs in previous works based on SDSS data,
we adopted `blueshift = 0.1' when using this code, and obtained
absolute magnitudes in z=0.1 blushifted bandpasses.

In LF calculations, we exploited two methods 
based on maximum likelihood approach.  One is the parametric maximum likelihood
method introduced by Sandage, Tammann \&
Yahil\ (\cite{1979apj...232..352S}, the so called STY method).  It
is based on the probability for a galaxy of redshift $z_i$ and
absolute magnitude $M_i$ to be included in a magnitude-limited
sample:

\begin{equation}
p_i=\frac{\phi(M_i)}{\int_{max[M_{min}(z_i),M_1]}^{min[M_{max}(z_i),M_2]}\phi(M)dM}
\end{equation}

where $M_{min}(z_i)$ and $M_{max}(z_i)$ are the minimum and maximum absolute magnitudes a galaxy at redshift $z_i$ can have in order to be included in the sample, $M_1$ and $M_2$
are the absolute magnitude limit of the sample.
In order to correct for the incompleteness, 
the following correcting factor $Fac_n(mag_{bin})$ is defined for every galaxy in each bin, 

\begin{equation}
Fac_n(mag_{bin})=\frac{1}{Fraction_{red}(mag_{bin})+Fraction_{orange}(mag_{bin})}
\end{equation} 
Here $Fraction_{red}$ and $Fraction_{orange}$ correspond to red bars and orange bars presented in Figure \ref{fig:completeness}. 
We assumed that the incompleteness depends only on the apparent magnitude.
A Schechter function (Schechter~\cite{1976apj...203..297}) for $\phi(M)$
was adopted when maximizing the log-likelihood function $ln\mathcal{L}$,

\begin{equation}
\phi(M)=0.4ln(10)\phi_*10^{-0.4(M-M_*)(\alpha+1)}exp(-10^{-0.4(M-M_*)})
\end{equation}

\begin{equation}
ln\mathcal{L}=\sum_{i}^{N_{gal}}{Fac_ilnp_i}   \label{L}
\end{equation}

The other method, the Stepwise Maximum Likelihood Method (SWML), is a
non-parametric method described by Efstathiou, Ellis \& Peterson\
(\cite{1988mnras...232..431E}).  This method does not depend on any
assumption on the particular form of an LF. The sample is divided
into $N_{bin}$ bins according to the absolute magnitude, and the LF
can be calculated as:

\begin{equation}
\phi(M)=\phi_{i}, M_{i}-\Delta/2<M<M_{i}+\Delta/2,i=1,2,...,N_{bin}
\end{equation}
where $\phi_{i}$ is the value of the LF in each bin, 
which can be derived iteratively by maximizing 
a log-likelihood function similar to that in Eq. (\ref{L}).

For either method, we used the minimum variance estimator (Davis \& Huchra\ (\cite{1982apj...254..437D})
to calculate independently the normalization constant $\bar{n}$ of each LF.
$\bar{n}$ represents the number density of galaxies, and it can be expressed as:

\begin{equation}
\phi(M)=\bar{n} \phi^{*}(M)
\end{equation}
where $\phi^{*}(M)$ is the unit-normalized luminosity function.

We did not carry out the correction for cosmic evolutionary effects 
because it may introduce significant
uncertainties due to our relatively small
sample size and large number of parameters involved in the calculation 
(Blanton et al.~\cite{2003apj...592..819B}).

In the calculation of errors of the STY LFs, we used the jackknife re-sampling method which has been used in many previous work (Blanton et al.~\cite{2003apj...592..819B}; 
Loveday et al.~\cite{2012mnras...420..1239L}).
We divided our total region into 8 sub-regions of approximately equal area, each time omitting 1 region in the calculation, and got a set of parameters $x^{k}=\{\alpha,M_{*},\bar{n}\}$.
The statistical variance of the fitting parameter $x^{k}=\{\alpha,M_{*},\bar{n}\}$ can be written as 

\begin{equation}
var(x^{k})=\frac{N-1}{N}\sum_{n=1}^{N}(x_{n}^{k}-\bar{x}^{k})^{2}
\end{equation}
where N=8 is the number of jackknife regions and $\bar{x}^{k}$ is the
mean of the parameter $x_{n}^{k}$ fitted while excluding region i. It
should be pointed out that, for large samples covering widely
separated sky areas (e.g. Blanton et al.~\cite{2003apj...592..819B}),
Jackknife method can include uncertainties due to
large-scale structure across the survey, namely the cosmic variance.
However, due to relatively small area of our survey, this does not
apply to our results. Therefore, while
the uncertainties of parameters ($\alpha$, $M_{*}$) in our work may be
underestimated, the cosmic variance is added to the error of $\bar{n}$.
It is
estimated according to (Peebles et al.~\cite{Peebles1980}; Somerville et al.~\cite{Somerville2004}; Xu et al.~\cite{Xu2012})

\begin{equation}
\sigma^{2}_{cv}=J_{2}(\gamma)\times(r_0/r_{sample})^{\gamma}
\end{equation}
where $r_{0}=5.59h^{-1}Mpc$ and $\gamma=1.84$ (Zehavi et al. \cite{Zehavi2005}) are parameters in the two point correlation function, $r_{sample}$ represents the radius of sample volume, and $J_{2}$ is 
\begin{equation}
J_{2}(\gamma)=\frac{72}{(3-\gamma)\times(4-\gamma)\times(6-\gamma)\times 2^{\gamma}} .
\end{equation}
In $^{0.1}r$ band, the cosmic variance contributes $\sim 63.3\%$ of the error of $\bar{n}$. 
   
For SWML LFs, the errors of $\phi_{i}$ were calculated using 
inversion of the information matrix as described in Efstathiou, Ellis \& Peterson\ (\cite{1988mnras...232..431E}).

In every band, we also calculated the luminosity density using parameters of
corresponding STY LF:
\begin{equation}
j=\int_{0}^{\infty} dLL\phi(L)=\phi_{*}L_{*}\Gamma(\alpha+2).
\end{equation}

\section{The results and discussion} \label{sec:results}

\subsection{LFs and luminosity densities} \label{sec:LFandLD}
\begin{figure*}
\centering
\includegraphics[bb=450 0 1000 450,width=0.35\textwidth]{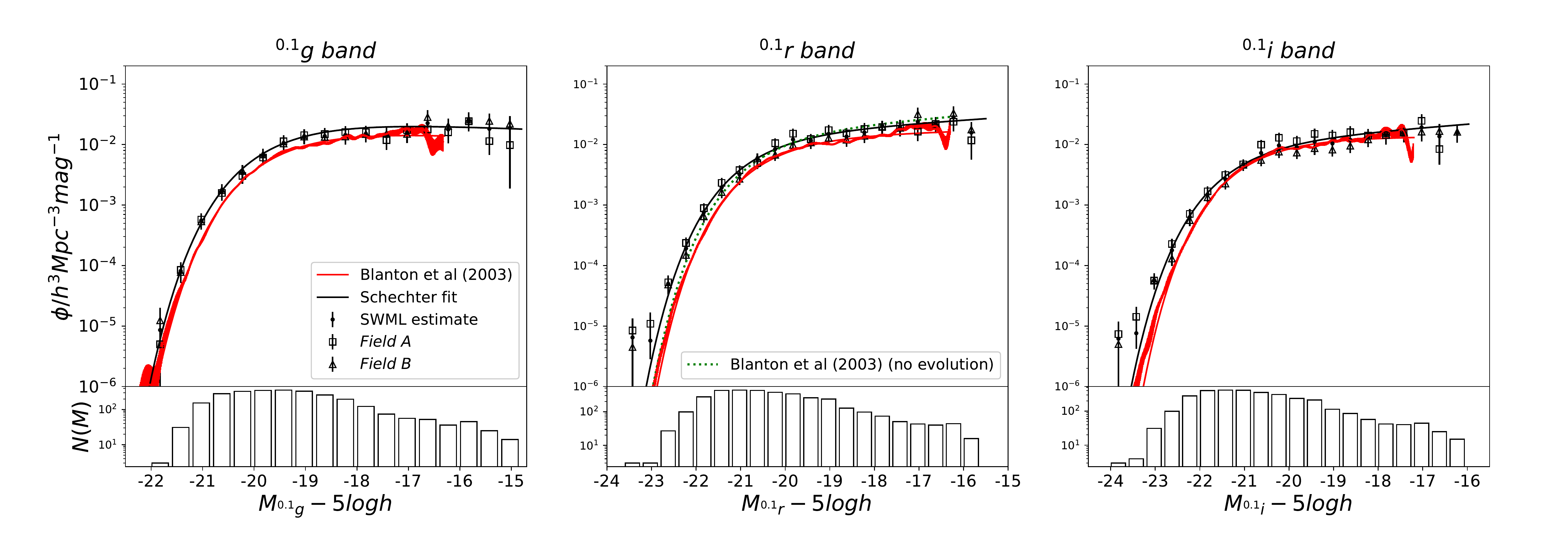}
\caption {Upper panel: the LFs of LaCoSSPAr in SDSS $^{0.1}g, ^{0.1}r, ^{0.1}i$ bands. The black dots are the SWML LFs with error bars and black lines are the results of best Schechter function fit by using STY method.
The red thick curves show the LFs of Blanton et al.\ (\cite{2003apj...592..819B}) along with its error region, and red lines are the best-fitting Schechter functions, both are under a simple evolution assumption.
LFs of Field A and Field B are also plotted as squares and triangles, respectively.
In r-band panel, we overplot the dotted line shown in Fig.15 in Blanton et al.\ (\cite{2003apj...592..819B}), which represent the best-fit Schechter function based on a sample of $\sim$ 10,000 galaxies, taking no account for the effect of evolution.
Lower panel: histograms of the number of galaxies in each bin for SDSS $^{0.1}g, ^{0.1}r, ^{0.1}i$ bands.
}
\label{fig:LF1}
\end{figure*}

As shown in Figure \ref{fig:LF1}, our LFs obtained using the
parametric STY method and the nonparametric SWML method agree well in
all three bands. In every band our LF extents approximately $\sim 1$
mag toward the fainter end compared to that of Blanton et al.\
(\cite{2003apj...592..819B}), because the LAMOST redshift survey is
deeper than SDSS. The marginally significant discrepancy with the
results of Blanton et al.\ (\cite{2003apj...592..819B}) is 
mainly due to the cosmic evolution correction carried out
by them but omitted in this work (see Section 4.1). Indeed, when
compared to their $^{0.1}r$ LF without evolution correction (green
dotted line in the mid-panel of Figure \ref{fig:LF1}), the discrepancy
is reduced remarkably: the difference is $< 1\sigma$ for any
Schechter function parameter except for $M_*$ ($2.5 \sigma$, Table 2).
 Another reason for the differences between LFs
of Blanton et al.\ (\cite{2003apj...592..819B}) and ours could be due
to the cosmic variance. Both Field A and Field B from which our sample
was selected are affected by clusters (see Section 2.1). 
In Figure \ref{fig:LF1},
overplotted are the LFs derived using subsamples of sources in the two
fields, separately. The difference between results from the two
fields is mainly in the faint end. In the bright end of $^{0.1}r$ LFs,
results of the total sample and of the subsamples in the two
fields are all slightly higher than the non-evolution LF of Blanton et
al.\ (\cite{2003apj...592..819B}). Nevertheless, as shown in Table 2,
the difference between values of our $^{0.1}r$ band 
density parameter $\bar{n}$ and that of the non-evolution model of
Blanton et al.\ (\cite{2003apj...592..819B}) is only 7\%, significantly
less than $1 \sigma$.
In this work, we used 0.4 absolute magnitude bin for the SWML
estimates to ensures that there are adequate  number of galaxies in 
each bin.  From the lower panel in Figure \ref{fig:LF1} it can be seen
that in $^{0.1}r$ band, there are $\sim10$ galaxies for the faintest
bin.  This to some extent makes our errors bars in SWML estimates seem
comparable to Blanton et al.\ (\cite{2003apj...592..819B})'s results
(Figure \ref{fig:LF1}), though our sample size is much smaller.

Table \ref{tab:table2} lists our best-fitting parameters, the luminosity densities, number densities and their $1\sigma$ uncertainties in $^{0.1}g, ^{0.1}r, ^{0.1}i$ bands. 
For comparison, we also list the parameters of the $^{0.1}r$ band non-evolution 
LF of Blanton et al.\ (\cite{2003apj...592..819B}).
The uncertainties of best fitting parameters in our work are larger than Blanton et al.\ (\cite{2003apj...592..819B}).
because small size samples selected from small sky areas are used in this work.
Our $^{0.1}r$ band luminosity density agrees very well with that of Blanton et al.\ (\cite{2003apj...592..819B}) based on the non-evolution LF. 
Our luminosity densities are also consistent with the luminosity density evolution trend shown in Fig. 20. of Loveday et al.\ (\cite{2012mnras...420..1239L}).

\begin{table}
\begin{center}
 \renewcommand\arraystretch{1.0}
 \caption{Three parameters $\phi_{*}, M_{*}, \alpha$ of Schechter function fits, luminosity densities and number densities
          in $^{0.1}g^{0.1}r^{0.1}i$ bands for this work and Blanton et al.\ (\cite{2003apj...592..819B}).}
 \label{tab:table2}
 \begin{tabular}{ccccccccccc} %
  \hline
  \multicolumn{11}{c}{This work} \\ 
  Band & {} & $\phi_{*}\ (10^{-2}h^{3}Mpc^{-3})$ & {} & $M_{*}-5log_{10}h$ & {} & $\alpha$ & {} & $j+2.5log_{10}h$\ (mag\ in\ $Mpc^{3}$) & {} & $\bar{n}\ (10^{-2}h^{3}Mpc^{-3})$ \\
  $^{0.1}g$ & {} & $2.93\pm0.54$ & {} & $-19.49\pm0.05$ & {} & $-0.91\pm0.08$ & {} & $-15.61\pm0.22$ & {} & $2.70\pm0.50$\\
  $^{0.1}r$ & {} & $1.65\pm0.36$ & {} & $-20.69\pm0.06$ & {} & $-1.12\pm0.08$ & {} & $-16.32\pm0.27$ & {} & $1.52\pm0.33$\\
  $^{0.1}i$ & {} & $1.22\pm0.24$ & {} & $-21.15\pm0.06$ & {} & $-1.14\pm0.06$ & {} & $-16.47\pm0.24$ & {} & $1.13\pm0.22$\\
  \multicolumn{11}{c}{Blanton et al.(2003) (no evolution)} \\
  Band & {} & $\phi_{*}\ (10^{-2}h^{3}Mpc^{-3})$ & {} & $M_{*}-5log_{10}h$ & {} & $\alpha$ & {} & $j+2.5log_{10}h$\ (mag\ in\ $Mpc^{3}$) & {} & $\bar{n}\ (10^{-2}h^{3}Mpc^{-3})$\\
  $^{0.1}r$ & {} & $1.77$ & {} & $-20.54$ & {} & $-1.15$ & {} & $-16.28$ & {} & 1.63\\
  \hline
 \end{tabular}
\end{center}
\end{table}

\begin{figure*}
\centering
\includegraphics[bb=0 0 450 450,width=0.45\textwidth]{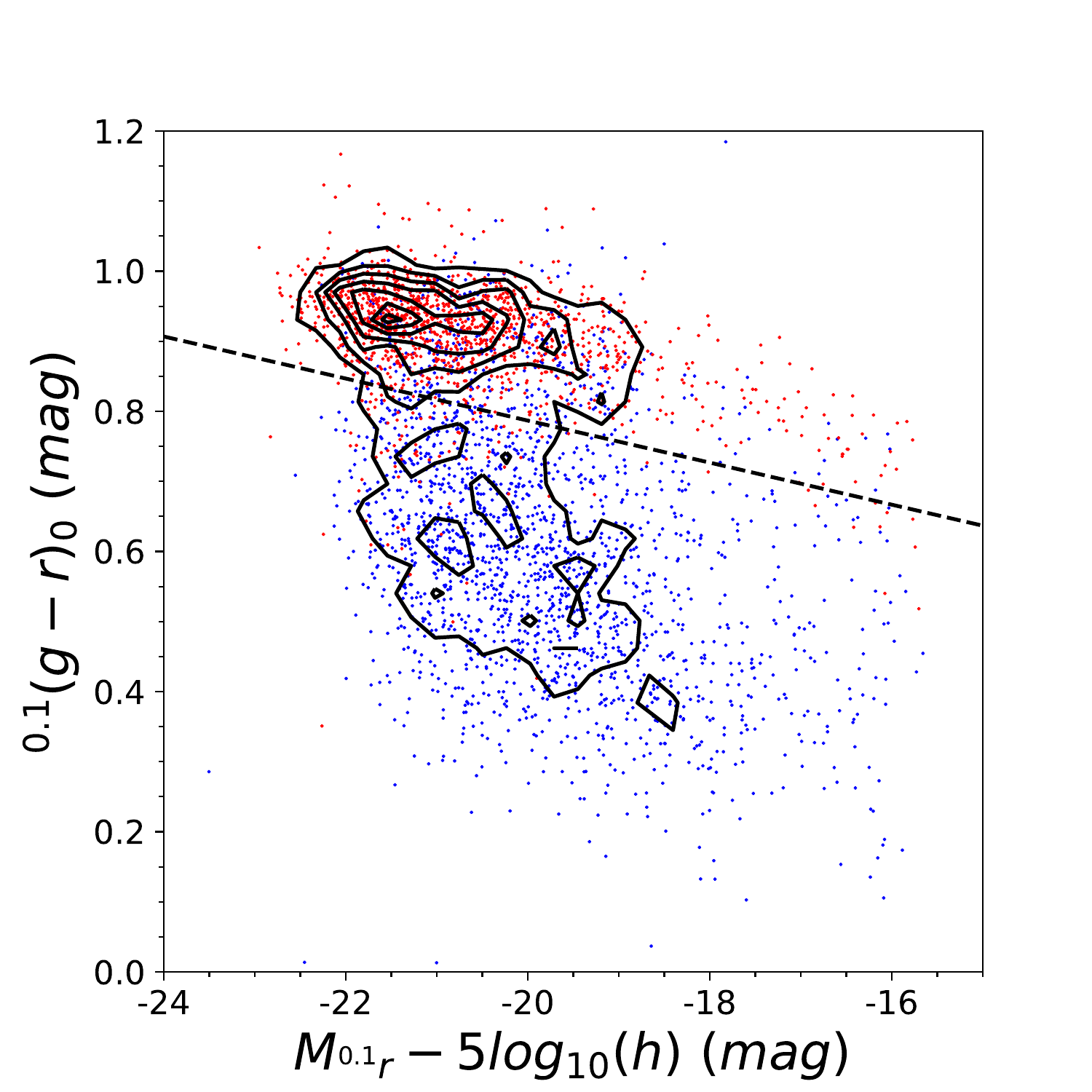}
\caption {The color-Magnitude diagram for the r-band subsample described in Section \ref{subsec:samselection}. 
Red and blue color dots are used to distinguish absorption line galaxies and emission line galaxies.
The black contour lines for r-band subsample and a black dashed separation line for the rest-frame bi-modality of galaxies (Zehavi et al.~\cite{Zehavi2011}) are also plotted.
}
\label{fig:c_M_relation}
\end{figure*}

\subsection{Dependence of LFs on spectral type}  \label{sec:LFon_spectype}

Depending on whether there are obvious emission lines in their spectra, 
Yang et al.\ (\cite{2018yang}) divided galaxies
observed in LaCoSSPAr survey into emission line galaxies and
absorption line galaxies.  The absorption
line galaxies sample comprises 1375 typical passive galaxies.
Figure \ref{fig:c_M_relation} presents the color-magnitude diagram of
$M_{^{0.1}r}$ vs. $^{0.1}(g-r)_{0}$ for the r-band subsample described in
Section \ref{subsec:samselection}.  Here $^{0.1}(g-r)_{0}$ is the
rest-frame color for 0.1 blushifted g and r band.  Red dots represent
the absorption line galaxies and blue dots the emission line
galaxies.  A contour diagram and a separation line are also
overplotted in Figure \ref{fig:c_M_relation}.  The color-magnitude
separation line (black dashed) is taken from Zehavi et
al. (\cite{Zehavi2011}):
\begin{equation}
^{0.1}(g-r)_{0}=0.21-0.03\times M_{^{0.1}r}
\end{equation} 

\begin{figure*}
\centering
\includegraphics[bb=450 0 900 450,width=0.35\textwidth]{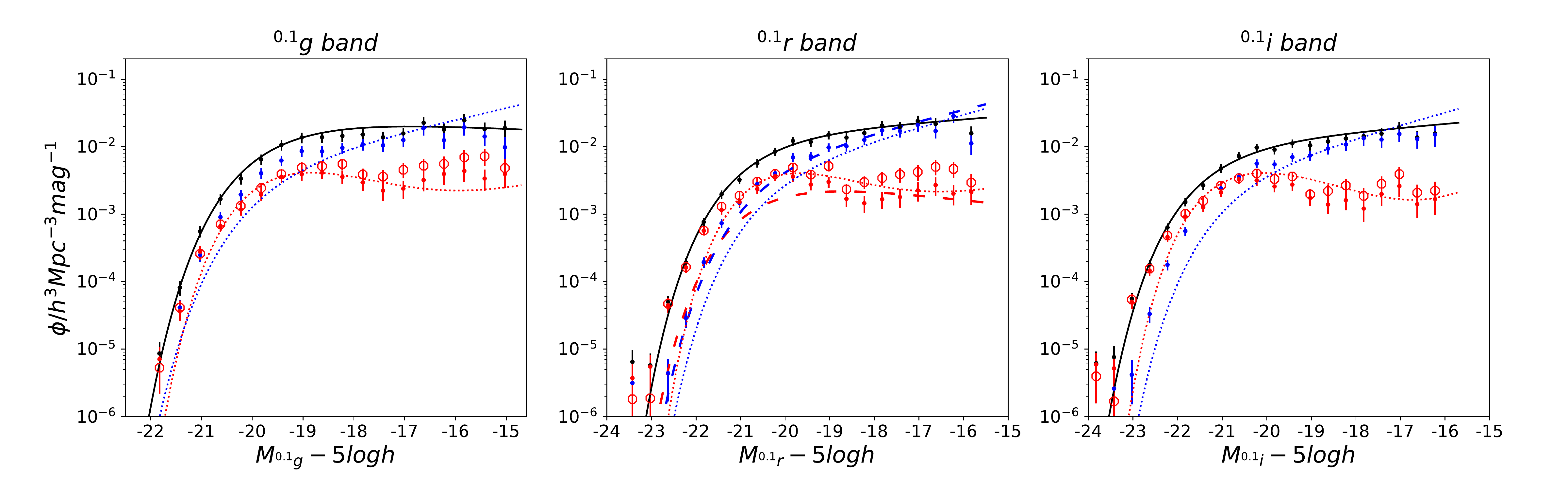}
\caption {SWML estimates of LFs for emission line galaxies (blue dots and error bars) and absorption line galaxies (red dots and error bars) in $^{0.1}g, ^{0.1}r, ^{0.1}i$ bands in our work.
          The red unfilled circles are the SWML estimates of LFs for red galaxies corresponding to galaxies located above the separation line in Figure \ref{fig:c_M_relation}.
          The black dots, error bars and black solid lines are the same as LFs presented in Figure \ref{fig:LF1}.
          The dashed lines plotted in $^{0.1}r$ band are the best fitting Schechter function of LFs for blue galaxies (blue dashed line) and red galaxies (red dashed line) estimated 
          by Montero-Dorta et al. (\cite{2009mnras...399..1106M}).
          The dotted lines in each band present the LFs for blue (blue dotted line) and red (red dotted line) galaxies at low redshift ($z<0.1$) 
          from Fig. 13 of Loveday et al. (\cite{2012mnras...420..1239L}), which are all fitted with the double-power-law Schechter function.
}
\label{fig:LFea}
\end{figure*}

In Figure \ref{fig:LFea}, SWML LFs of SDSS $^{0.1}g, ^{0.1}r,
^{0.1}i$-band for emission line galaxies (blue dots), absorption line
galaxies (red dots), and red galaxies (those located above the
separation line in Figure \ref{fig:c_M_relation}, red open circle) are
plotted, respectively, and are compared to the LFs of the total
samples. The absorption line galaxies show higher number densities
than emission line galaxies at the luminous end ($Mag_{^{0.1}r}$ or
$Mag_{^{0.1}i}$ $>21.5$ mag) in $^{0.1}r$ and $^{0.1}i$ bands.  In
each band, the LF of emission line galaxies appears to have a
Schechter function profile with a steeper faint end slope than that of
the total sample.  As for absorption line galaxies (and red galaxies),
the LFs show an obvious dip at $M-5log_{10}h\sim -18.5$ mag in all three
bands.  A standard Schechter function cannot provide a good fit to
the LF of absorption line galaxies over entire magnitude range.

Similar results have been found in many previous works 
on LFs of passive galaxies in 
different photometric bands and different redshift ranges
(Madgwick et al. \cite{2002mnras...333..133M};\ Wolf et
al. \cite{Wolf2003};\ Blanton et al. \cite{Blanton2005};\ Salimbeni et
al. \cite{Salimbeni2008}; \ Loveday et
al. \cite{2012mnras...420..1239L};\ L$\acute{o}$pez-Sanjuan et
al. \cite{Lopez2017}).  Madgwick et al.\
(\cite{2002mnras...333..133M}) investigated galaxy luminosity
functions for the 2dF survey in $M_{b_{j}}$ band for different
spectral types.  They divided their galaxies into four spectral types
by introducing a new parameter $\eta$, which identifies the average
emission and absorption line strength in the galaxy rest-frame
spectrum.  Their LF for `Type 1' galaxies (absorption line galaxies)
shows an obvious dip at $M_{b_{j}}-5log_{10}h\sim -16$ mag. 

For comparisons, in Figure \ref{fig:LFea} we overplot the LFs of
red (red dotted lines) and blue galaxies (blue dotted lines) by
Loveday et al. (\cite{2012mnras...420..1239L}) in three bands, and by
Montero-Dorta \& Prada\ (\cite{2009mnras...399..1106M}) in $^{0.1}r$\
band (red and blue dashed line).  The LFs of blue galaxies of Loveday
et al. (\cite{2012mnras...420..1239L}) are in general lower than those
of our emission line galaxies.  A possible cause for this, beside the
difference in definitions of blue galaxies and emission line
galaxies, could be the cosmic evolutionary effect because our galaxies
have a higher median redshift ($\bar{z}=0.104$) than theirs (all with
$z<0.1$) and we did not do any evolutionary correction. For red
galaxies, Loveday et al. (\cite{2012mnras...420..1239L}) fitted the
LFs with double-power-law Schechter functions, in the form of
\begin{equation}
\phi(M)=0.4ln(10)exp[-10^{-0.4(M-M_{*,S})}] \times \\
\{\phi_{1,S}10^{-0.4(M-M_{*,S})(\alpha_{1,S}+1)}+ \\
\phi_{2,S}10^{-0.4(M-M_{*,S})(\alpha_{2,S}+1)}\}.
\end{equation}
They show poor
agreements with our results for both absorption line galaxies and red
galaxies.  Montero-Dorta \& Prada\ (\cite{2009mnras...399..1106M})
used the Schechter function to fit their LFs for red and blue
galaxies. Their $^{0.1}r$\ band LF of blue galaxies shows a much
better agreement with ours than Loveday et
al. (\cite{2012mnras...420..1239L}), but the LF of red galaxies is
significantly different from ours.

We found that a double-Gaussian function, 
as defined in what follows, can provide significantly better fits
to the LFs of absorption line galaxies and red galaxies:
\begin{equation}
\phi=\phi_{1,G}exp(-\frac{(M-M_{1,G})^{2}}{2\sigma^{2}_{1,G}})+ 
\phi_{2,G}exp(-\frac{(M-M_{2,G})^{2}}{2\sigma^{2}_{2,G}})
\end{equation} 

\begin{figure*}
\centering
\includegraphics[bb=450 0 900 450,width=0.35\textwidth]{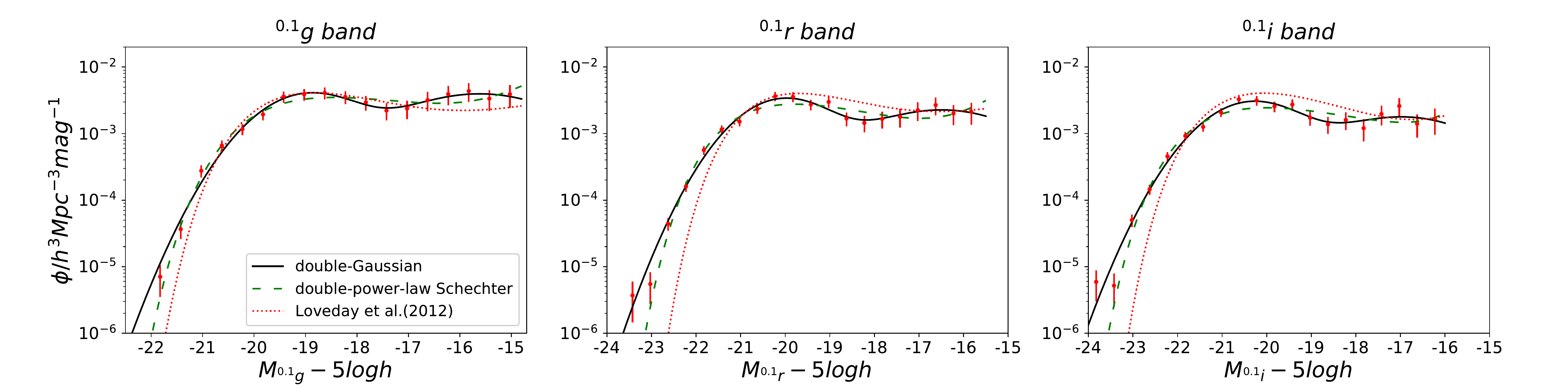}
\caption {Double-Gaussian function (black solid lines) and double-power-law Schechter function (green dashed lines) fit to LFs of absorption line galaxies in $^{0.1}g, ^{0.1}r, ^{0.1}i$ bands.
          The red dots are SWML estimates of absorption line galaxies showed in Figure \ref{fig:LFea}.
          Red dotted lines represent the LFs for red galaxies at low redshift ($z<0.1$) from Fig. 13 of Loveday et al. (\cite{2012mnras...420..1239L}).          
}
\label{fig:LFagalfit}
\end{figure*}

In Figure \ref{fig:LFagalfit} and Tables 3-4 we compare results of
double-Gaussian fittings and double-power-law Schechter function
fittings. For absorption line galaxies, the former not only
provides a much better fit to the dip at $M_{^{0.1}r}-5log_{10}h\sim -18.5$ mag,
but also results in smaller reduced-$\chi^{2}$'s in all three bands than the
latter.  While the double-power-law Schechter function has one
characteristic absolute magnitude ($M_{*,S}$), the double-Gaussian
function has two characteristic absolute magnitudes $M_{1,G}$ and
$M_{2,G}$.  This may hint at a bi-modality in the population of
absorption line galaxies, with the two sub-populations having
distinctively different characteristic luminosities (masses):  
The more massive sub-population has the
luminosity of $L^*$ galaxies, while galaxies in the less massive
sub-population are $\sim 3.5$ mag (i.e. $\sim 25\times$) fainter. 

Peng et al. (2010; 2012) argued that passive galaxies are mainly
formed through two distinct processes of ``mass quenching" and
``environment quenching''.  The massive central galaxies
(characterized by $L^*$ galaxies) are presumably quenched by the first
process, and low mass satellite galaxies are quenched by the second
process. Is the ``bi-modality'' of the absorption line galaxies
consistent with this theory? To answer this question,
we carried out the following test: Firstly we cross-matched
our sample of absorption line galaxies with the SDSS-DR7 based
NYU-VAGC catalog that Yang et al. (2007) used for group
identifications, and then checked the matches for memberships in Yang's
groups.  After excluding galaxies associated with one-galaxy groups
(i.e. single galaxies) and with groups having no halo mass estimates
(uncertain groups), we found 83 absorption line galaxies (70 bright
galaxies with $M_{^{0.1}r}-5log_{10}h < -18.5$, 13 faint galaxies with
$M_{^{0.1}r}-5log_{10}h > -18.5$) belonging to 30 groups. Among the 70
bright galaxies (``more massive galaxies''), 26 (37\%) are the
brightest or the most massive galaxies in their groups, and another
7 (10\%) are the second brightest galaxies in groups with three 
or more members, suggesting that $\sim 50\%$ of these galaxies are the
master galaxies in groups. On the other
hand, none of the faint galaxies is the brightest neither the most
massive galaxy in any group that they belong. Actually, 8 out of
the 13 faint absorption line galaxies belong to a single rich group
(group-ID 280, with 34 identified members), and indeed they appear to
be the ``satellite'' galaxies in the group, all with the $M_{r}$ rank
after the 20th. Our results seem to agree with the hypotheses
that the bright and massive absorption line galaxies tend to be master
galaxies in groups, while most faint and less massive absorption line
galaxies are satellites, consistent with the theory of Peng
et al. (2010; 2012). It is worth noting that, because of the
poor coverage of the SDSS spectroscopic survey in our fields
(Figure~3), only a small fraction of the absorption line galaxies have
matches in the NYU-VAGC catalog. Also, the number of faint galaxies
(83) is much less than that of bright galaxies (1292) in the absorption line galaxy sample since the finding volume of
the former is much smaller than that of the latter.

\begin{table*}
\begin{center}
 \renewcommand\arraystretch{1.0}
 \caption{Least squares fitting parameters of double-Gaussian function to LFs of absorption line galaxies in each band.}
 \label{tab:table3}
 \begin{tabular}{cccccccc}
  \hline
  Band & $\phi_{1,G}$ & $M_{1,G}-5log_{10}h$ & $\sigma_{1,G}^{2}$ & $\phi_{2,G}$ & $M_{2,G}-5log_{10}h$ & $\sigma_{2,G}^{2}$ & $\chi^{2}_{\nu}$ \\
  {}   & $(10^{-2}h^{3}Mpc^{-3})$ & {} & {} & {$(10^{-2}h^{3}Mpc^{-3})$} & {} & {} &{} \\
  \hline
  $^{0.1}g$ & 0.40 & -18.95 & 1.36 & 0.40 & -15.63 & 3.66 & 0.66\\
  $^{0.1}r$ & 0.34 & -20.04 & 1.53 & 0.23 & -16.50 & 4.40 & 0.93\\
  $^{0.1}i$ & 0.30 & -20.35 & 1.66 & 0.18 & -16.96 & 4.40 & 1.07\\
  \hline
 \end{tabular}
\end{center}
\end{table*}

\begin{table*}
\begin{center}
 \renewcommand\arraystretch{1.0}
 \caption{Least squares fitting parameters of double-power-law Schechter function to LFs of absorption line galaxies in each band.}
 \label{tab:table4}
 \begin{tabular}{ccccccc}
  \hline
  Band & $M_{*,S}-5log_{10}h$ & $\alpha_{1,S}$ & $\alpha_{2,S}$ & $\phi_{1,S}$ & $\phi_{2,S}$ & $\chi^{2}_{\nu}$ \\
  {}   & {} & {} & {} & $(10^{-5}h^{3}Mpc^{-3})$ & $(10^{-2}h^{3}Mpc^{-3})$ &{} \\
  \hline
  $^{0.1}g$ & -19.48 & -1.90 & -0.53 & 9.05 & 0.79 & 0.80\\
  $^{0.1}r$ & -20.68 & -2.03 & -0.52 & 1.91 & 0.66 & 1.31\\
  $^{0.1}i$ & -21.16 & -2.57 & -0.64 & 0.08 & 0.53 & 1.67\\
  \hline
 \end{tabular}
\end{center}
\end{table*}

\section{Summary} \label{sec:summary}

LAMOST is one of the most powerful telescopes in accessing the spectra
of celestial objects. As a key project of LAMOST, LaCoSSPAr provides
the most complete dataset of LAMOST ExtraGAlactic Survey (LEGAS) up to
now.  In this work, we analyzed the redshift incompleteness in
LaCoSSPAr survey quantitatively, and obtained the first measurements
of the galaxy LFs in the $^{0.1}g$, $^{0.1}r$, and $^{0.1}i$ bands
using LAMOST spectroscopic data.

We used both parametric (STY) and non-parametric (SWML) maximum
likelihood methods to construct LFs, and found good agreements between
the results.  Our LFs are compatible to previous works using SDSS
data.  Thanks to deeper magnitude limit of LAMOST, compared to results
based on SDSS data, we were able to extend the faint end of the LFs by
$\sim 1$ mag.  Our luminosity densities are consistent with the
luminosity density evolution obtained by Loveday et al.\
(\cite{2012mnras...420..1239L}).
 
We divided our sample into emission line galaxies and absorption line
galaxies, and derived their LFs separately. Our results show that, in
every band, the SWML estimate of emission line galaxies LF has a
Schechter function profile with a steeper faint end slope than that of
the total sample. The LFs of absorption line galaxies show an obvious
dip near $\sim 18.5$ mag in all three bands, and cannot be fitted by
Schechter functions. On the other hand double-Gaussian functions, with
two characteristic absolute magnitudes $M_{1,G}$ and $M_{2,G}$,
provide excellent fits to them. This may hint at a bi-modality in the
population of absorption line galaxies (representing passive
galaxies), with the two sub-populations having distinctively different
characteristic luminosities (masses): The more massive sub-population
has the luminosity of $L^*$ galaxies, while galaxies in the less
massive sub-population are $\sim 3.5$ mag (i.e. $\sim 25\times$)
fainter. Investigations using the group catalog of Yang et al (2007)
indicate that the former tend to be the master galaxies in groups
while most of the latter are satellites.

This work is based on a small size galaxy sample within $\sim 40$ $deg^{2}$ survey area which leads to large statistic uncertainties in LF estimates.
In the future, we can expect a large-area covering sample when LAMOST finishes its LEGAS survey which can give us a better-constrained and unbiased estimates for LFs.

\begin{acknowledgements}
We would like to thank the staff of LAMOST at Xinglong Station for their excellent support during our observing runs.

This project is supported by 
the National Key R\&D Program of China (No. 2017YFA0402704);
the National Natural Science Foundation of China (Grant No.11733006, U1531245); 
the National Science Foundation for Young Scientists of China (Grant No.11603058); 
the Guoshoujing Telescope Spectroscopic Survey Key Projects.
CKX acknowledges support by the National Natural Science Foundation of China No. Y811251N01. His work is sponsored in part by the Chinese Academy of Sciences (CAS), through a grant to the CAS South America Center for Astronomy (CASSACA) in Santiago, Chile.

Guoshoujing Telescope (the Large Sky Area Multi-Object Fiber Spectroscopic Telescope LAMOST) is a National Major Scientific Project built by the Chinese Academy of Sciences. Funding for the project has been provided by the National Development and Reform Commission. 
LAMOST is operated and managed by the National Astronomical Observatories, Chinese Academy of Sciences.

Funding for SDSS-III has been provided by the Alfred P. Sloan Foundation, the Participating Institutions, the National Science Foundation, and the U.S. Department of Energy Office of Science. The SDSS-III web site is http://www.sdss3.org/.

\end{acknowledgements}

\label{lastpage}

\end{document}